%% file: M87_chevron.tex
\documentclass[letter]{aa}
\usepackage{graphicx}
\usepackage{txfonts}
\usepackage{aalongtable}
 \usepackage{newclude}
\usepackage[round]{natbib}
\usepackage{color}
\bibpunct{(}{)}{;}{a}{}{,}
\graphicspath{{./images/}}
\begin{document}

\title{The build-up of the cD halo of M87 - evidence for \\ 
accretion in the last Gyr \thanks{Based on observations made with the VLT at Paranal
    Observatory under programs 088.B-0288(A) and 093.B-066(A), and
    with the SUBARU Telescope under program S10A-039.}}

\author{A. Longobardi\inst{1}, M. Arnaboldi\inst{2},
  O. Gerhard\inst{1}, J.C. Mihos\inst{3}}

\offprints{A. Longobardi}

\institute{Max-Planck-Institut f\"ur Extraterrestrische Physik,
  Giessenbachstrasse, D-85741 Garching, Germany \\
  e-mail: alongobardi@mpe.mpg.de, gerhard@mpe.mpg.de
  \and European Southern Observatory, Karl-Schwarzschild-Strasse 2,
  D-85748 Garching, Germany \\
  e-mail: marnabol@eso.org
  \and Department of Physics and Astronomy, Case Western University, 10900 Euclid Avenue, Cleveland, OH 44106, USA\\
  e-mail: mihos@case.edu}

\date{in press, A\&A}

   \authorrunning{A. Longobardi et al.}
   \titlerunning{Accretion in the halo M87}

% \abstract{}{}{}{}{} 
% 5 {} token are mandatory
 
\abstract 
% context heading ({optional) 
{} 
{We present kinematic and photometric evidence for an accretion
event in the halo of the cD galaxy M87 in the last Gyr.}
{Using velocities for $\sim 300$ planetary nebulas (PNs) in the M87
  halo, we identify a chevron-like substructure in the PN phase-space.
  We implement a probabilistic Gaussian mixture model to identify PNs
  that belong to the chevron. From analysis of deep V-band images of
  M87, we find that the region with the highest density of chevron PNs
  is a crown-shaped substructure in the light.}
{We assign a total of $\mathrm{N_{PN,sub}}=54$ to the substructure,
  which extends over $\sim$50 kpc along the major axis where we also
  observe radial variations of the ellipticity profile and a colour
  gradient. The substructure has highest surface brightness in a 20kpc
  $\times$ 60kpc region around 70 kpc in radius. In this region, it
  causes an increase in surface brightness by $\gtrsim$60\%. The
  accretion event is consistent with a progenitor galaxy with a V-band
  luminosity of $\mathrm{L=2.8\pm1.0\times 10^{9} L_{\odot,\mathrm{V}}}$, a
  colour of $\mathrm{(B-V)=0.76\pm0.05}$, and a stellar mass of $\mathrm{M=
    6.4\pm2.3\times 10^{9} M_{\odot}}$.}
{The accretion of this progenitor galaxy has caused an important
  modification of the outer halo of M87 in the last Gyr.  By itself it
  is strong evidence that the galaxy's cD halo is growing through the
  accretion of smaller galaxies as predicted by hierarchical galaxy
  evolution models.}


   \maketitle
% 
% ________________________________________________________________

 \include*{Introduction}

\include*{Section_2}

 \include*{Section_3}

 \include*{Section_4}

 \include*{Section_5}

\begin{acknowledgements}
  AL is grateful to J. Elliott for helpful comments which improved the
  manuscript. JCM is supported by NSF grant 1108964.
\end{acknowledgements}

\bibliographystyle{aa}
\bibliography{PNrefs}
\newpage
\newpage
\end{document}

%% file: Introduction.tex
\section{Introduction}
\label{sec1}
According to the current theory of hierarchical formation and
evolution of structures, accretion events are believed to have an
important role in the cosmological build up of stellar halos in
elliptical galaxies \citep{delucia07}, responsible for their
growth at relatively low redshifts \citep[$z < 2$;][]{oser10}.  In
dense environments accretion is even more dramatic, such that close to
the dynamical centre of the cluster, central cluster galaxies are
expected to have the majority of their stars accreted
\citep{laporte13,cooper14}.

On the observational front, the outer stellar envelopes are observed
to increase in mass by a factor of $\sim4$ since $z=2$
\citep{vandokkum10}. The presence of an accreted component is
usually identified as an excess of light over the extrapolation of the
galaxy's inner profile \citep{zibetti05,gonzales07,dsouza14}, or by high
Sersic indexes \citep[$n>4$;][]{kormendy09}. Observations of blue
colour gradients from the centres of early-type galaxies towards their
outskirts \citep{peletier90,liu05,rudick10}, mainly attributed to a
gradient in metallicty \citep{tamura00,loubser12,montes14} are also in
agreement with a change in stellar properties driven by the accretion
of smaller systems. Records of accretion events are also revealed
in the form of spatially extended low-surface brightness features in
the outskirt of stellar halos \citep{mihos05,vandokkum14,duc15}. These
\textit{substructures} are not in a phase-mixed equilibrium in the host
galaxy potential and, therefore, can in principle be traced as
kinematic features in velocity phase-space.

M87 is one of the nearest central galaxies \citep[at an adopted
distance of 14.5 Mpc;][]{ciardullo02,longobardi15}, close to
the dynamical centre of the Virgo cluster~\citep{binggeli87}. It is
considered a type-cD galaxy \citep{weil97,kormendy09} with an extended
stellar envelope that reaches $\sim150$ kpc in radius. A blue colour
gradient towards its outskirts \citep{rudick10} has been interpreted
as age and metallicity gradients \citep{liu05,montes14}, consistent
with a late build-up of its halo. The orbital properties of
  globular clusters (GCs) \citep{agnello14} and ultra compact dwarfs
  \citep{zhang15} also favour accretion onto the halo. M87 has been
the target of several deep imaging surveys \citep{mihos05,rudick10},
and its close proximity has made it possible to identify hundreds of
planetary nebulas (PNs), and to measure their line-of-sight velocities
\citep[LOSVs;][]{doherty09,longobardi15}.

In this letter we use the synergy between PN kinematics and deep
imaging to identify an on-going accretion event in the outer
halo of M87, which we find to be a non-negligible perturbation of
the galaxy properties at the distances where it is traced.

%% file: Section_2.tex
\begin{figure*} \centering
\includegraphics[width=7.1cm]{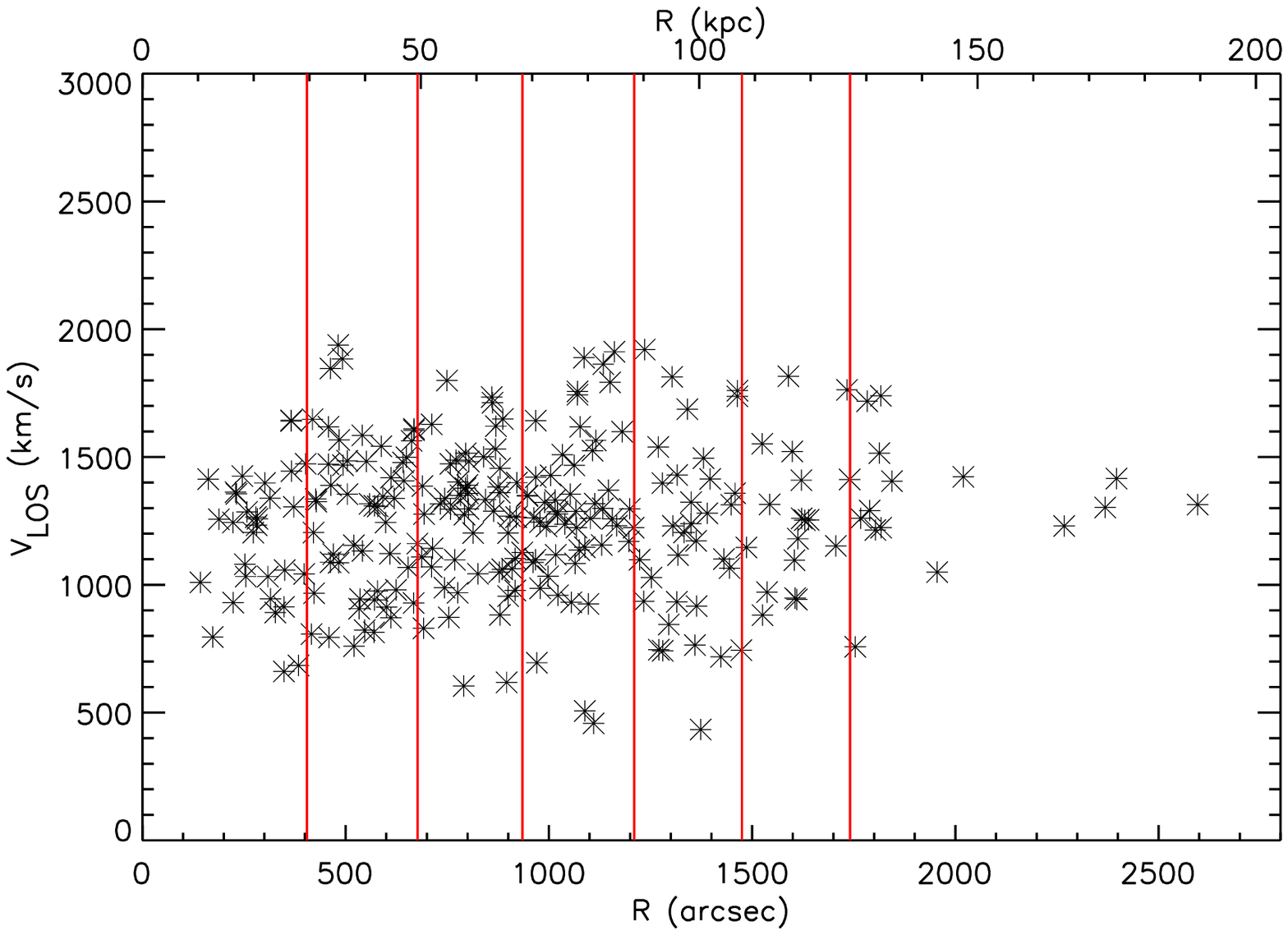}
\includegraphics[width=7.6cm, clip=true, trim=0cm 6.5cm 1.2cm 6.5cm]{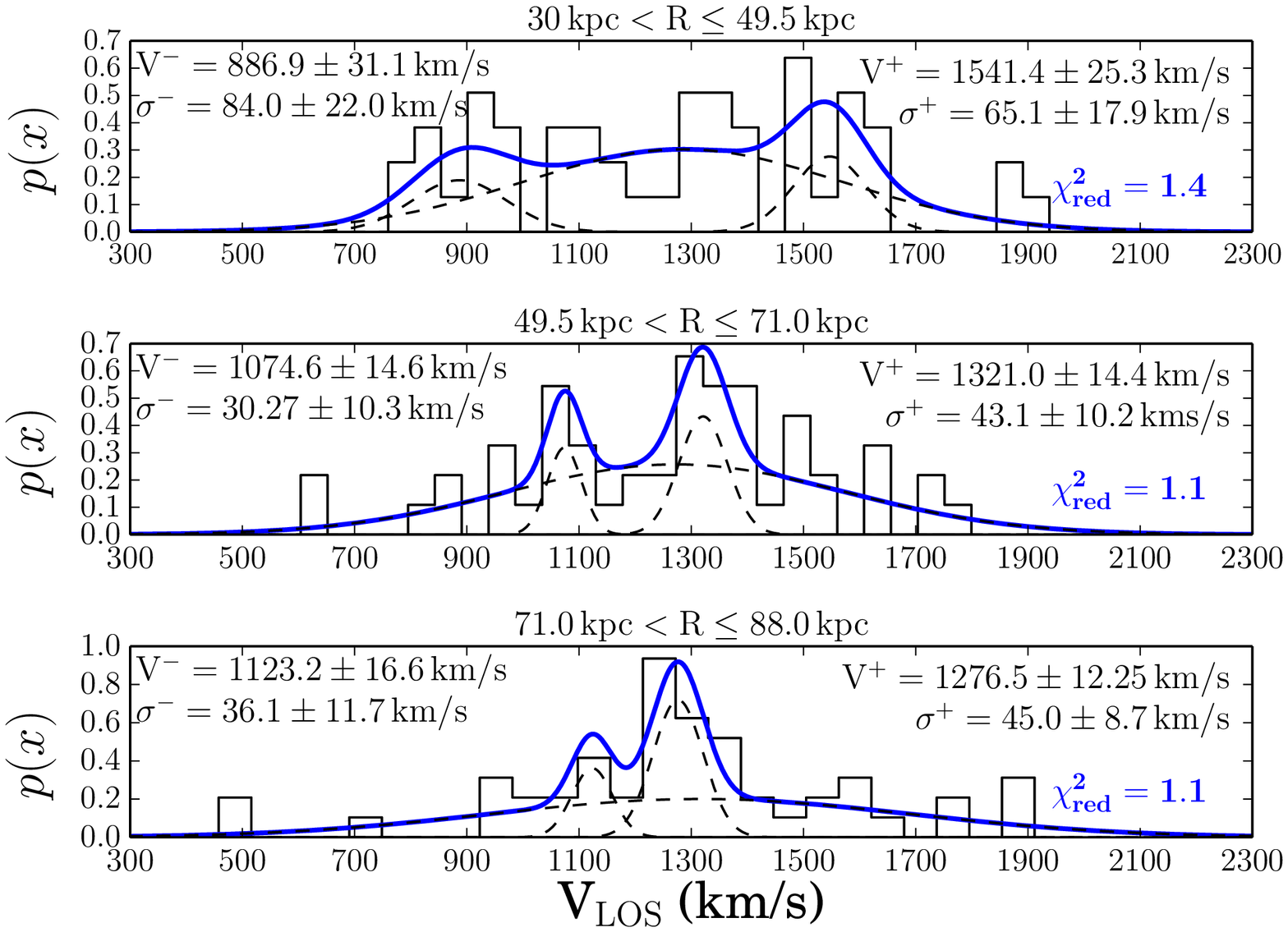}
\includegraphics[width=7.1cm]{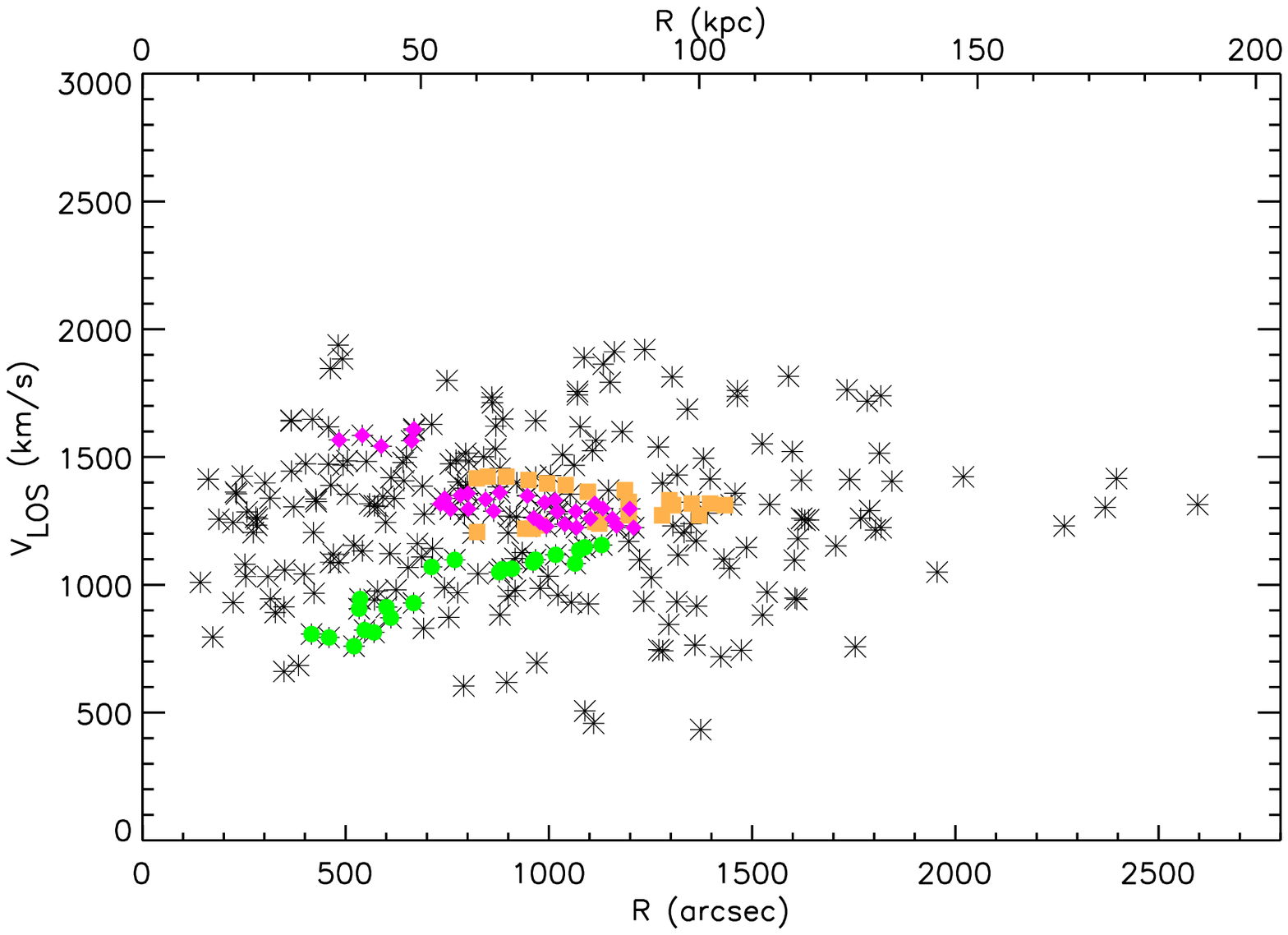}
\includegraphics[width=7.6cm, clip=true, trim=0cm 6.5cm 1.2cm 6.5cm]{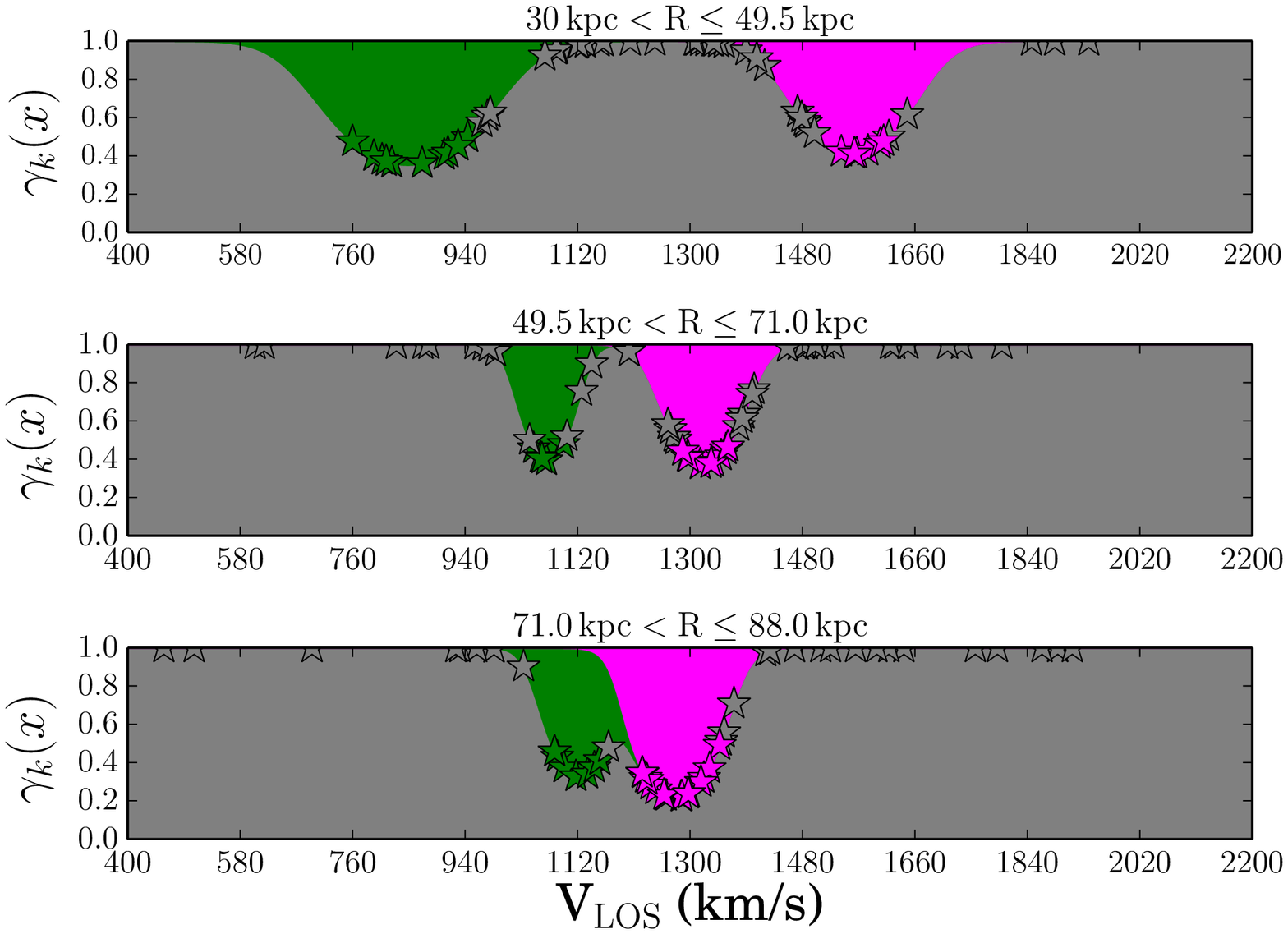}
\vspace{-0.3cm}
\caption{{\bf Top-left}: Projected phase-space,
  $\mathrm{V}_{\mathrm{LOS}}$ vs. major axis distance ($R$), for all
  spectroscopically confirmed PNs (black asterisks) in the halo of
  M87. Red lines separate the elliptical annuli analysed to isolate
  cold components associated to the substructure. {\bf Top-right}:
  Histograms of the LOSVD in the three elliptical annuli. In each
  panel, the blue lines show the best-fit model computed as a
  combination of three Gaussians. Black-dashed lines show the relative
  contribution of each component to the LOSVD, with parameters for the
  cold components given in the plot. {\bf Bottom-left}: As in the
  top-left panel, however, the green circles and magenta diamonds show
  the PNs associated with the cold secondary peaks in the
  LOSVD. Orange squares show a kinematically selected GC substructure
  from \citet{romanowsky12}.  {\bf Bottom-right}: Probability that a
  PN is drawn from the halo component (dark grey area) or from the
  chevron (green, magenta areas). Stars represent PN probabilities at
  their measured V$_{\mathrm{LOS}}$.}\label{GMM_chevron}
%Coloured
%  stars represent chevron PNs and probabilities at their measured
%  V$_{\mathrm{LOS}}$.}\label{GMM_chevron}
\end{figure*}

\section{Kinematic evidence for an accretion event in M87}
\label{sec2}

We acquired kinematic data from the FLAMES/VLT spectroscopic survey
for a large sample of PNs in the outer regions of M87
\citep{longobardi13,longobardi15}. The total sample consists of 254
objects classified as M87 halo PNs and 44 intracluster PNs, for which
we have obtained LOSVs with an estimated median velocity accuracy of
$\rm 4.2\, kms^{-1}$ . We concentrate here on the M87 halo PNs that
cover a range of radii from $\sim15-150$ kpc.

We see a notable chevron (or ``V'' shape) structure in the projected
phase-space of the PN sample as shown in Fig.~\ref{GMM_chevron} (top
left). To isolate this kinematical substructure we utilise a
three-component Gaussian mixture model to identify high-density,
narrow features on top of a broader distribution. We note that there
is not enough data to statistically favour this model over simpler
models  with BIC or AIC \citep[Bayesian/Akaike Information
  Criteria;][]{liddle07}; however, it is visually indicated and we
will confirm it with photometry in Sec.\ref{sec4}. A brief description
of the technique is given in the following paragraph; for more details
we refer the reader to \citet{pedregosa11}.

A Gaussian Mixture Model (GMM) is a probabilistic model, which assumes
that a distribution of points can be described as a linear combination
of $K$ independent Gaussian probability density functions (PDFs), or
components, expressed by:
\begin{equation}
p( x)=\sum_{k=1}^{K} p_{k}(x\, |\, \mu_{k}, \sigma_{k})P_{k},
\end{equation} 
\label{sec3}
where, $x$ is a data vector (here the LOSVs), $P_{k}$ is the mixture
weight that satisfies the conditions $0 \le P_{k} \le 1$ and
$\sum_{k=1}^{K}P_{k}$=1, and $ p(x\, |\, \mu_{k}, \sigma_{k})$ are the
individual Gaussian PDFs, with mean $\mu_{k}$, and dispersion
$\sigma_{k}$. The GMM classifier implements the
Expectation-Maximization (EM) algorithm, i.e. an iterative process
that continuously updates the PDF parameters until convergence is
reached. At the end of the EM procedure, the posterior probabilities,
$\gamma_{k}(x)$ for a data value to belong to each of the $k$ Gaussian
components are returned. These are described by:
\begin{equation}
\gamma_{k}(x)=\frac{p_{k}(x\, |\, \mu_{k}, \sigma_{k})P_{k}}{p(x)}.
\label{post_p}
\end{equation}
To apply the GMM to our LOSV distribution (LOVSD) we bin the PN M87
halo sample into seven elliptical annuli, or stripes in
phase-space, covering the entire PN velocity phase-space. The
LOSVD in each annulus is analysed as a combination of three Gaussians,
where the centres ($\mu_{k}$), widths ($\sigma_{k}$), and weights
($P_{k}$) are treated as free parameters in the EM algorithm, and have
uncertainties $\sigma_{\mu_{k}}=\sigma_{k}/\sqrt{S}$,
$\sigma_{\sigma_{k}}=\sigma_{k}/\sqrt{2S}$, and
$\sigma_{P_{k}}=P_{k}/\sqrt{S}$, with $S=[\sum_{n}\gamma_{k}(x_{n})]$
\citep{mackay03}.

We find cold components in three out of seven elliptical bins, for
which we show the histogram of the data, along with the best-fit GMM
and reduced $\chi ^2$ in Fig.~\ref{GMM_chevron} (top-right panel). We
also plot (bottom-left panel) the LOSV phase-space for the 254 PNs in
the halo of M87: black crosses represent PNs of the smooth halo LOSVD
of M87, while magenta diamonds and green dots are PNs that have a
higher probability (see eq.~\ref{post_p}) to belong to the
chevron. Finally, we show the probability that a given PN is drawn
from each of the components as a function of its velocity
(bottom-right panel).  The GMM assigned a total of 54 PNs to the
chevron substructure, which covers 700$\arcsec$ ($\sim$ 50 kpc) for
major axis distances $500\arcsec<R<1200\arcsec$. The separation,
$\Delta \mathrm{V}$, between the two peaks of the cold components
becomes smaller at larger distances. For the three elliptical bins, it
is $\Delta \mathrm{V}=~654.5~\pm~40.1,\, 246.4~\pm~20.5,\,\rm and\,
153.3~\pm~20.6\, kms^{-1}$, respectively.  At $R\sim1200\arcsec$
($\sim90$ kpc) the width of the chevron goes to zero with LOSVs close
to the galaxy's systemic velocity \citep[$\mathrm{V_{sys}=1275\,
  kms^{-1}}$;][]{longobardi15}\footnote{ The stability of the fitted
  parameters and the measured distance of the chevron edge were tested
  with 100 GMM runs for different mock data sets and initialisation
  values.}. PNs on the arms of the chevron are seen on both the
northern and southern sides of the galaxy as is shown in
Fig.~\ref{Image_sub} (see Sect.~\ref{sec3}). The broad Gaussian with
average mean velocity $\sim1290\, \rm{kms^{-1}}$ and dispersion
$\sim320\, \rm{kms^{-1}}$ in the three bins traces the M87 halo.

The search for kinematic features in the phase-space of GCs has
resulted in the discovery of a similar chevron-like structure
\citep{romanowsky12}, shown in Fig.\ref{GMM_chevron} above as orange
squares. Though the morphology in the phase-space is similar it
  differs in a number of physical properties: the width goes to zero
at $R_{\mathrm{GC}}\sim1500 \arcsec$ with
$\mathrm{V}_{\mathrm{LOS,GC}}=1307\, \rm km s^{-1}$, versus $R\sim1200
\arcsec$ and $\mathrm{V}_{\mathrm{LOS}}\sim1250~\pm21\, \rm km s^{-1}$
for the PNs. Moreover, the 27 chevron GCs show a very different
  spatial distribution with the highest density of points on the NE
photometric minor axis \citep{romanowsky12,dabrusco15}, and few GCs
near the crown substructure traced by the PNs.

%% file: Section_3.tex
\section{Localising the substructure with deep imaging}
\label{sec3}
\begin{figure*} \centering
    \includegraphics[width=12.cm, clip=true, trim=0.9cm 0cm 0.5cm 0.cm]{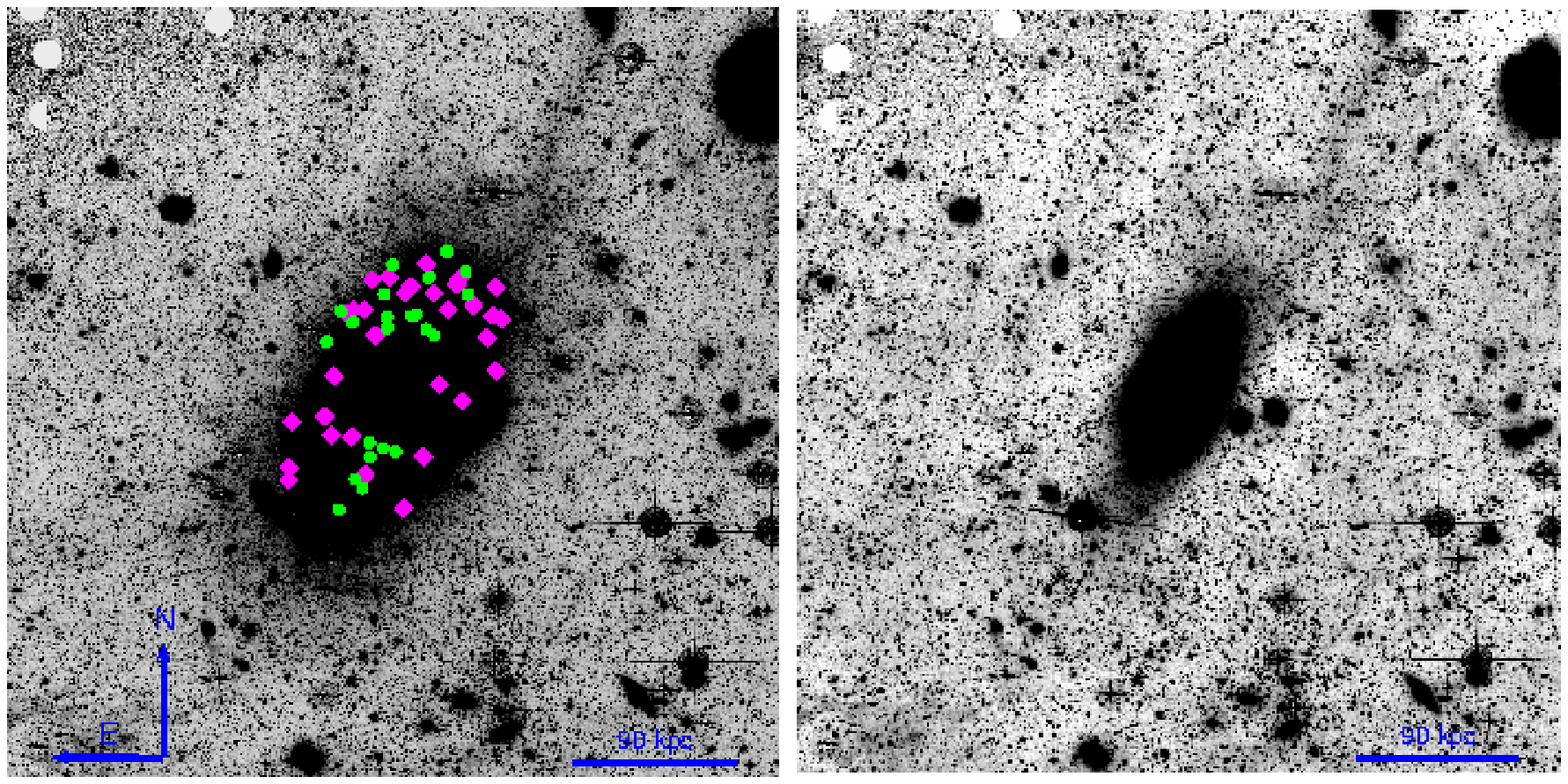}
\hspace{-0.1cm}
    \includegraphics[width=5.9cm, clip=true, trim=0cm 0.cm 0.2cm 0.cm]{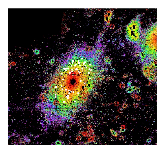}
  
    \vspace{-0.3cm}
   \caption{Spatial and colour distribution associated with the
     kinematic substructure identified in the phase-space of the M87
     halo PNs. {\bf Left panel}: V-band image of a $1.6 \times 1.6$
     deg$^2$ centred on M87 from~\citet{mihos15}. Full circles, and diamonds
     indicate the spatial position of the M87 halo PNs in the chevron
     substructure. Magenta and green colours indicate PN LOSVs above and
     below $\mathrm{V_{LOS}} = 1254\, \rm km s^{-1}$, the LOSV at the
     end of the chevron. {\bf Central panel}: Unsharp masked image
     of M87 median binned to enhance faint structures. The
     crown-shaped substructure is visible at distance of
     800\arcsec-1200\arcsec ($\sim60-90$ kpc) along the major axis, NW
     of M87. Details are given in Section~\ref{sec3}. The blue line
     measures 90 kpc. {\bf Right panel}: (B-V) colour image of M87
     from~\citet{mihos15} with chevron PNs overplotted (white
     dots). The dashed ellipse indicates the isophote at a major axis
     distance of 1200\arcsec. The crown is found in a region where the
     (B-V) colour is on average 0.8, bluer than on the minor
     axis.}\label{Image_sub}
\end{figure*}
In Fig.~\ref{Image_sub} (left), we show the position of the chevron
PNs overplotted on $1.6 \times 1.6$ deg$^{2}$ V-band image of M87,
with an estimated surface brightness limit of $\mathrm{\mu_{V}=28.5\,
  magarcsec^{-2}}$ \citep{mihos15}.  Here we see the large spatial
extent of the substructure associated with the chevron PNs. Because of
its shape on the image we will refer to it as the {\sl veil of
  M87}. Now we are interested to see if this feature is also visible
in the optical light.

To this end, we constructed an unsharped masked image, which is the
difference between the original image and a smoothed image. We
utilised the IRAF task $\textbf{fmedian}$ to smooth the original image
by using a window with a size of $1450\arcsec \times
1450\arcsec$. This window size was chosen so that it contained the
large scale extension of the substructure. By looking at the highest
concentration of PNs, in the NW region of
M87, it can be seen to extend over many hundreds of arcseconds
($\sim 800 \arcsec$). Thus, the adopted box size is $\sim
1.8$ times larger than the long side of the feature.

The results of the unsharped masking can be seen in
Fig.~\ref{Image_sub} (central panel), where the high frequency
structures are now clearly visible. A previously unknown debris
structure, with a crown-like shape, can be seen on top of M87 at the
NW side.  This we refer to as the {\sl crown} of M87's veil.  It has a
characteristic width of $\sim300\arcsec$, an extension of
$\sim800\arcsec$, and is almost perpendicular to M87's photometric
major axis. %~\citep[P.~A.$\sim-25.6\degr$;][]{kormendy09}.
Just as the PN spatial distribution showed, the edge of this feature
is found at $R\sim1200\arcsec\, (\sim 90\, \mathrm{kpc})$.

Over the same major axis distances at which the substructure is
located, we also observe variations in the M87 ellipticity
profile. Between $300\arcsec <R< 800\arcsec$ the ellipticity increases
and then flattens to a value of $e\sim0.43$ for $R>800\arcsec$
\citep{kormendy09}.  In particular, the region at which the gradient
flattens reflects the crown-like overdensity.

%% file: Section_4.tex
\section{Physical properties of the accreted satellite} 
\label{sec4}
\subsection{Luminosity and $\alpha$-parameter}
To understand the physical origin of this structure, we compute its
total luminosity. We do it in a region with size
$\sim800\arcsec\times300\arcsec$, after the subtraction of the local
background, determined using a region photometry method \citep[for
  more details see][]{rudick10}. We find a total luminosity in the
range of $\mathrm{L_{crown}}= 3.7\pm0.9 \times
10^{8}\mathrm{L_{\odot,V}}$. When compared to the luminosity
determined in the same region from the original image, $\mathrm{L}\sim
5.9 \times 10^{8}\mathrm{L_{\odot,V}}$, it is clear that at these
distances the substructure represents a significant fraction,
$\sim60\%$, of the total light.

In the region of the crown we count $\mathrm{N_{PN,crown}}=12\pm3$
PNs, while we find a total of $\mathrm{N_{PN,chevron}}=54\pm7$ PNs
associated to the entire chevron (see Sect.~\ref{sec2}). By correcting
these numbers for incompleteness factors as in \citet{longobardi15}
these become 19 and 142, respectively.  Hence, by scaling
$\mathrm{L_{crown}}$ to the total number of PNs associated to the
chevron we obtain the total luminosity associated to the progenitor of
the M87 veil to be $\mathrm{L}\sim 2.8\pm1.0 \times
10^{9}\mathrm{L_{\odot,V}}$.

The total number of PNs is proportional to the total bolometric
luminosity of the parent stellar population, and the proportionality
is quantified with the luminosity-specific PN density, or,
$\alpha-$parameter \citep{buzzoni06}. Utilising the computed
luminosity, and the completeness-corrected estimate for
$\mathrm{N_{PN,chevron}}$, we can calculate the $\alpha$-parameter for
the progenitor of the substructure. Considering that the typical
probability of the $\mathrm{N_{PN,chevron}}$ PN to belong to the
chevron is $\sim0.7$ (Fig.~\ref{GMM_chevron}), we obtain
$\alpha_{2.5}=1.8\pm0.7 \times 10^{-8}$N$_{\mathrm{PN}}$L$^{-1}_{\odot,\mathrm{bol}}$. 
Here $\alpha_{2.5}$ is 2.5 mag down the luminosity function as in
\citet{longobardi15}, and we have assumed a bolometric correction for
the V-band of BC$_{\mathrm{V}}$=0.85 \citep{buzzoni06} and BC$_{\odot}$=-0.07 for
the Sun.

\subsection{Colour and Mass}
In Fig.~\ref{Image_sub} (right panel) we show the chevron PNs
overplotted on the B-V colour image of M87, that combines the V-band
data (see Sect.~\ref{sec3}) with deep B-band photometry with a surface
brightness limit of $\mathrm{\mu_{B}=29\, mag\, arcsec^{-2}}$. It is
interesting to notice that close to $R\sim 1200\arcsec$ the colour
shows an azimuthal variance, such that along the photometric minor
axis the measured values are redder \citep{mihos15}. This feature
correlates with the spatial number density of the chevron PNs, showing
a deficit in number along the photometric minor axis. This suggests
that the bluer regions are the result of the accreted material on top
the light from M87's halo. In particular, the crown structure is
measured to have integrated colour $\mathrm{(B-V)}= 0.76\pm0.05$.

(B-V) colour is a good estimator of the mass-to-light-ratio,
$\Upsilon^{\star}$, of the underlying stellar population. By adopting
$\Upsilon^{\star}_{\mathrm{V}} = 2.3$ for (B-V)=0.76
\citep{mcgaugh14}, the total stellar mass associated to the disrupted
galaxy is then $\mathrm{M=6.4\pm2.3 \times 10^{9}M_{\odot}}$.

From the distribution and velocities of chevron PNs in
Fig.~\ref{Image_sub} a possible interpretation of the satellite orbit
could be that it was first disrupted entering M87 from the South
(along the green dots), with the debris then moving up North, turning
around in the crown region, and coming back down on both sides across
M87 (the veil, magenta dots). The velocities would then imply that the
northern side of M87 is closer to the observer. The dynamical
time for such an orbit is of order $\lesssim 1$ Gyr \citep{weil97}.

 %The quenching of star formation as consequence of gas stripping from
 %the accreted satellite results in a reddening of the tidal debris
 %with respect the progenitor of $\Delta \mathrm{(B-V)}=0.2$
 %\citep{feldmann08}, leading to progenitor with (B-V)=0.5. From this
 %value and the computed total luminosity we can get an estimate of the
 %total stellar mass to be $ \mathrm{M_{prog}\sim2.9 \times
 %  10^{9}M_{\odot}}$, having assumed a mass-to-light ratio
 %$\Upsilon^{\star}_{\mathrm{V}} = 1.4$ \citet{mcgaugh14}.

%% file: Section_5.tex
\section{Summary and Conclusion} 
\label{sec5}

In this letter we have presented kinematic and photometric evidence
for an accretion event in the halo of the cD galaxy M87. This event is
traced by PNs whose velocity phase-space shows a distinct chevron-like
feature, which is a result of the incomplete phase-space mixing of a
disrupted galaxy.  At major axis distances of $R\sim60-90\, \rm kpc$,
where the width of the chevron goes to zero, a deep optical image
shows the presence of a crown-like substructure that contributes
$\gtrsim 60\%$ of the total light in this area.

The \textit{crown} of M87's {\sl veil} is the densest part of the entire
substructure, which covers $\sim50\, \rm kpc$ along the
major axis. In this region also a radial variation in M87's
ellipticity profile is observed. Looking at the spatial distribution
of all the chevron PNs, it traces the azimuthal variation observed in
the colour of M87, showing a deficit in number of tracers along the
photometric minor axis where the galaxy is redder, and a higher
fraction where the substructure is strongest and the colour is bluer.

We determined several physical properties of the disrupted satellite:
a total luminosity of $\mathrm{L=2.8\pm1.0\times 10^{9}
  L_{\odot,\mathrm{V}}}$, colour (B-V)=$0.76\pm0.05$, and total
stellar mass of $\mathrm{M=6.4\pm2.3\times 10^9 \, M_{\odot}}$. The
inferred value for the $\alpha$-parameter is $\alpha=1.8\pm0.7 \times
10^{-8}$ N$_{\mathrm{PN}}$L$^{-1}_{\odot,\mathrm{bol}}$. The similar
colours of the accreted satellite and ICL suggest that the cD halo of
M87 is presently growing by the accretion of similar star-forming
systems as those that originate the diffuse IC component.

The evidence for on-going accretion in the outer halo of M87 is
consistent with the observed size growth of giant elliptical galaxies
and with predictions by theory.  The presence of the newly discovered
substructure within the halo of M87 demonstrates, that beyond a
distance of $\sim 60\, \rm kpc$, its halo is still assembling.